# Visual Exploratory Data Analysis of the Covid-19 Pandemic in Nigeria: Two Years after the Outbreak

Ugochukwu .E. Orji
Dept. of Computer Science
University of Nigeria, Nsukka
Enugu, Nigeria
ugochukwu.orji.pg00609@unn.edu.ng

Elochukwu Ukwandu
Dept. of Applied Computing & Engineering,
Cardiff School of Technologies,
Cardiff Metropolitan University, UK
eaukwandu@cardiffmet.ac.uk

Ezugwu .A. Obianuju
Dept. of Computer Science
University of Nigeria, Nsukka
Enugu, Nigeria
assumpta.ezugwu@unn.edu.ng

Modesta .E. Ezema
Dept. of Computer Science
University of Nigeria, Nsukka
Enugu, Nigeria
modesta.ezema@unn.edu.ng

Chikaodili .H. Ugwuishiwu
Dept. of Computer Science
University of Nigeria, Nsukka
Enugu, Nigeria
chikodili.ugwuishiwu@unn.edu.ng

Malachi .C. Egbugha
Dept of Computer Engineering,
Ahmadu Bello University, Zaria
malachi.egbugha3@gmail.com

*Abstract* — The outbreak of the coronavirus disease in Nigeria and all over the world in 2019/2020 caused havoc on the world's economy and put a strain on global healthcare facilities and personnel. It also threw up many opportunities to improve processes using artificial intelligence techniques like big data analytics and business intelligence. The need to speedily make decisions that could have far-reaching effects is prompting the boom in data analytics which is achieved via exploratory data analysis (EDA) to see trends, patterns, and relationships in the data. Today, big data analytics is revolutionizing processes and helping improve productivity and decision-making capabilities in all aspects of life. The large amount of heterogeneous and, in most cases, opaque data now available has made it possible for researchers and businesses of all sizes to effectively deploy data analytics to gain action-oriented insights into various problems in real time. In this paper, we deployed Microsoft Excel and Python to perform EDA of the covid-19 pandemic data in Nigeria and presented our results via visualizations and a dashboard using Tableau. The dataset is from the Nigeria Centre for Disease Control (NCDC) recorded between February 28th, 2020, and July 19th, 2022. This paper aims to follow the data and visually show the trends over the past 2 years and also show the powerful capabilities of these data analytics tools and techniques. Furthermore, our findings contribute to the current literature on Covid-19 research by showcasing how the virus has progressed in Nigeria over time and the insights thus far.

*Keywords— Data Analytics, Visual EDA, Microsoft Excel, Pandas, Tableau Dashboard, Covid-19.*

## I. INTRODUCTION

The World Health Organization (WHO) was officially alerted of the emergence of cases of pneumonia-like diseases in Wuhan city, China, on December 31st, 2019. By the end of January 2020, over 9,000 cases of the coronavirus were confirmed throughout China, with 213 deaths [1]. The disease has since become a worldwide emergency, as its rapid spread and high mortality rate in most western countries and parts of Europe have caused severe disruptions and unprecedented lockdowns all over the world. Patients with COVID-19 have been known to present with pneumonia, severe symptoms of acute respiratory distress syndrome (ARDS), and multiple organ failure [2, 3, 4, 5].

In Nigeria, NCDC is the public health institution in charge of the nation's response to infectious disease outbreaks and public health emergencies and is thus equipped by law to guide the nation in handling the coronavirus disease. Through the Presidential Steering Committee and the Federal Ministry of Health, the Federal Government of Nigeria activated an NCDC-led national COVID-19 Emergency Operations Centre (EOC) to coordinate the national public health response to the pandemic. The committee is also tasked with routinely providing updates and guidelines to the Nigerian public. This is also replicated at the state level, where the response activities are coordinated through Public Health EOCs in each State [6].

Visual EDA as an integrated approach to data visualization and analysis has evolved with the development of scientific methods to data visualization, emphasizing analytical reasoning through an interactive visual interface. Visual analytic methods provide analysts with a platform to combine their creativity and domain knowledge with the enormous storage and processing capacities of today's computers to gain insight into complex problems [7]. Visual analytics benefits largely from information retrieval methodologies, data management & knowledge representation, and data mining [8].

Organizations and businesses of all sizes increasingly utilize data visualization tools and skills because it is much easier to identify patterns and trends through a visual summary of information than by looking through thousands of rows on a spreadsheet. According to [9], the human brain is more susceptible to visual information, and since the aim of any data analysis is to gain insights, the result of data analysis is much more valuable when visualized. Furthermore, visualization makes it easier to communicate data analysis findings to stakeholders without creating confusion [10]. The numerous types of data visualizations available include; Line charts, Box plots, Scatter plots, Bar charts, Pie charts, Heat maps, Bar charts, Histograms, Network diagrams, etc.

Concretely, the contribution of this study is as follows:

1. Visually show the covid-19 trend in Nigeria between February 28th, 2020, and July 19th, 2022.

2. Demonstrate the effectiveness of state-of-the-art big data analytic techniques and tools.

3. Identify other useful insights from the data.

## II. LITERATURE REVIEW

In this section, we review some of the related works done by researchers using EDA in various fields to show trends and insights.

In the past few years, there have been increasing success by researchers in using EDA to generate insights and hypotheses on various problems in various fields. The authors in [11] showed how social media and crowdsourcing create a new external financing option for entrepreneurs. Their paper investigated the hashtag #crowdfunding on Twitter through the lens of social network theory. They performed EDA on a dataset of 2,732,144 tweets published during a calendar year to generate insights and hypotheses on the #crowdfunding users. Their results proved successful as EDA helped identify actionable information for decision-makers.

Today, data analytics has become an essential technique employed by most industries because of its ability to uncover hidden patterns in raw data and its forecasting capacity. Visual EDA tools like Tableau makes it possible to visually manipulate large amounts of data with ease as demonstrated by the research of [12] where EDA was deployed on a heart disease dataset to predict risk factors that cause heart diseases. The authors made a predictive analysis using the K-means clustering algorithm along with Tableau for visualization.

At the beginning of the covid-19 outbreak, there were a lot of uncertainties and panic, fearmongering, and misinformation. People sought information about the virus, but there was limited information available at the time. Thankfully, Johns Hopkins University kept a daily updated dataset of covid-19 cases from all over the world, which helped researchers and analysts study the outbreak and proffer possible solutions. In their paper, [13] presented a visual EDA of the covid-19 outbreak in countries based on the number of confirmed & recovered cases, and deaths recorded. The authors also did a comparative analysis of the mortality and recovery rate. They used K-means clustering to classify the countries according to the number of confirmed cases and deaths associated with the virus.

Likewise, the authors in [14] performed an EDA of the COVID-19 Open Research Dataset (CORD-19) in their effort to investigate the virus and its impact on people with Intellectual Disabilities (ID). They used a text mining technique that combined frequency analysis (frequency-inverse) and K-means clustering to identify full-text articles related to ID care in the CORD-19 dataset. Their method identified 259 articles related to ID in the dataset.

Finally, in a related work by [15], the authors used a linear regression technique to analyze covid-19 data reported worldwide from January to the end of August 2020. Their research observed the contagious nature of the virus but also indicated that the virus is relatively non-lethal as the data showed a lower universal death record compared to the recovered cases. Thus, they concluded that early test and diagnosis of covid-19 will eliminate critical/severe cases and reduce deaths.

**Research gap**

After reviewing the related works done in this area, we established that the use of data analytics techniques and tools for EDA is on the rise and the results achieved have been promising. We also observed that few papers have explored the trends and possible insights from the covid-19 data in Nigeria which was well covered in this study. Table 1 below shows the concepts covered by authors so far in this area.

Table 1: Comparison of similar studies

| Paper | Concept handled | This paper |
|---|---|---|
| [14] | Used a combination of text mining and K-means clustering to perform EDA and identify full-text articles related to ID care in the CORD-19 dataset. | Visual EDA of Covid-19 disease in Nigeria with Microsoft Excel, Python and Tableau. |
| [11] | Performed Network analytics using the Gephi open graph visualization platform. | |
| [12] | Used Tableau and K means clustering for EDA | |
| [13] | Used K-means clustering and python visualization libraries for EDA | |
| [15] | Performed EDA of worldwide covid-19 data using linear regression technique | |

## III. METHODOLOGY

This section describes our dataset, the techniques for preprocessing and cleaning the data, and the Tableau visualization software used to tell the data story. Fig. 1 shows a flowchart of the visual EDA process.

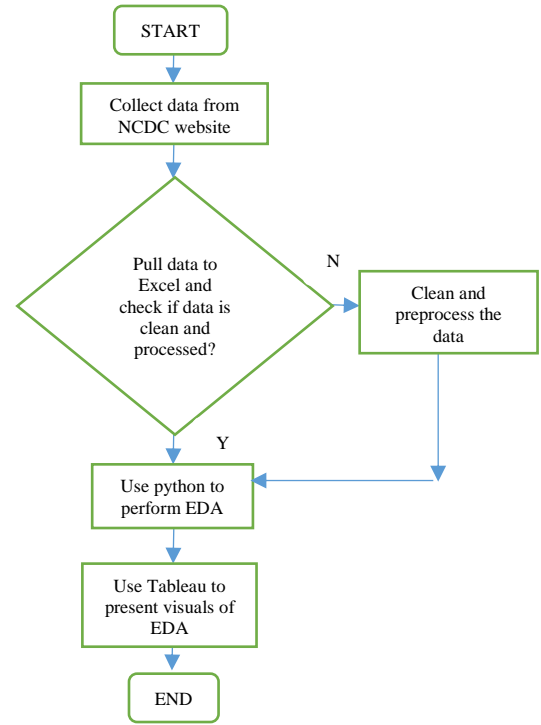

Fig. 1: flowchart of EDA process

**Dataset Description**

In this research, we used the NCDC dataset, which is publically accessible online [6]. The dataset comprises records of Covid-19 cases in all states of Nigeria between February 28th, 2020, and July 19th, 2022. The original dataset features include; State, Confirmed cases, Active cases, Discharged, and Deaths. We modified the dataset to include the Latitude and Longitude of the states for better visualization. In table 2 we provide a brief description of the data features.

Table 2: Brief description of the dataset

| Variable | Description | Data Types |
|---|---|---|
| State | All states in Nigeria plus FCT | object |
| Latitude | Geographical coordinates of states | float64 |
| Longitude | Geographical coordinates of states | float64 |
| Confirmed cases | Number of Lab confirmed cases | int64 |
| Active cases | Number of patients on admission | int64 |
| Discharged | Number of recovered and discharged patients | int64 |
| Deaths | Number of deaths associated with covid-19 | int64 |

**Data Preprocessing and Cleaning using Microsoft Excel and Python**

According to [16], checking data for errors and anomalies is the first task in analyzing any data set. Microsoft Excel is designed with some very impressive data-cleaning functions; making it a very important tool for initial data analysis. In most cases, the dataset from both online and offline sources is not clean, thus requiring further cleaning to ensure the data meets the required standard.

A major part of Excel Data Cleaning involves;

- Removing blank spaces,
- Deleting incorrect and outdated information
- Removing duplicates
- Detecting and deleting any unnecessary formatting
- Performing spelling checks
- Changing case to Lower/Upper/Proper options
- Deleting columns and rows you don't need
- Shorten text where necessary
- Correcting dates etc. [17].

The next step was examining the data for insights, relationships, and trends i.e EDA. Python is one of the most widely used EDA tools because it has numerous libraries to help manipulate the data to see the little details easily and not get lost in the data.

The tasks completed with Python in this project include;

- Check and confirm that the data types are in the correct structure
- Check and handle missing values via imputation and other techniques
- Perform a statistical summary of the data
- Find the pairwise correlation of all columns in the dataframe and check for outliers.

**Data Visualization using Tableau**

Tools like Tableau and Power BI are revolutionizing the development of scientific information visualization. The emphasis is on analytical reasoning using the interactive tools available on these visualization platforms [8]. Tableau has a big advantage as it allows users to visually manipulate the data and see hidden insights, patterns, and relationships between various variables in real time. You can connect various data sources in Tableau, and the most outstanding part is that you don't need to have any programming experience to use Tableau. The interactive pick-and-drop interface of Tableau makes it very user-friendly and easily adaptable.

Tableau also makes it easy for users to explore various visual views of the data. All users need is to choose a basic format or chart, then add more dimensions to the visualizations through the use of colors, shapes, and sizes for each variable. Furthermore, users can proceed to create a dashboard or story feature to produce a polished overview of the visualizations via Tableau's annotation and filtering tools [18].

### IV.  RESULT

In performing the EDA on our dataset, some vital questions were answered, including;

1. **What is the total number of confirmed cases of Covid-19 in Nigeria so far?**

As of July 19th, 2022, there are 259,007 confirmed cases of the Covid-19 virus in Nigeria. The data shows that Lagos, FCT, and Rivers state are the top three (3) states with the most confirmed cases, while Kebbi, Zamfara, and Kogi state are the three (3) states with the least confirmed cases of Covid-19 in Nigeria.

Fig. 2 below shows a breakdown of the confirmed cases in the 36 states and FCT:

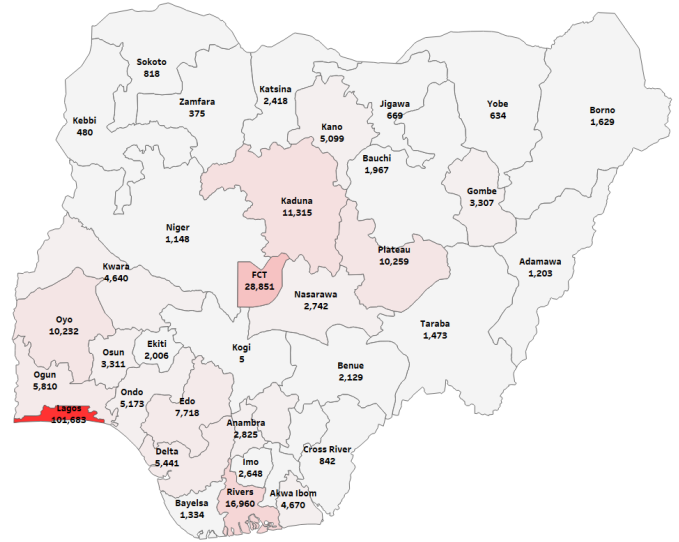

Fig. 2: Total confirmed cases for various states in Nigeria

**2. How many of these cases are currently active?**

As of July 19th, 2022, there are 5,378 active cases of the Covid-19 virus in Nigeria. The data shows that Lagos, Kwara, and Nasarawa states have the top three (3) number of active cases, while Katsina, Sokoto, Zamfara, and Kogi had zero (0) active cases of Covid-19. Fig. 3 below shows a breakdown of the active cases across the 36 states and FCT:

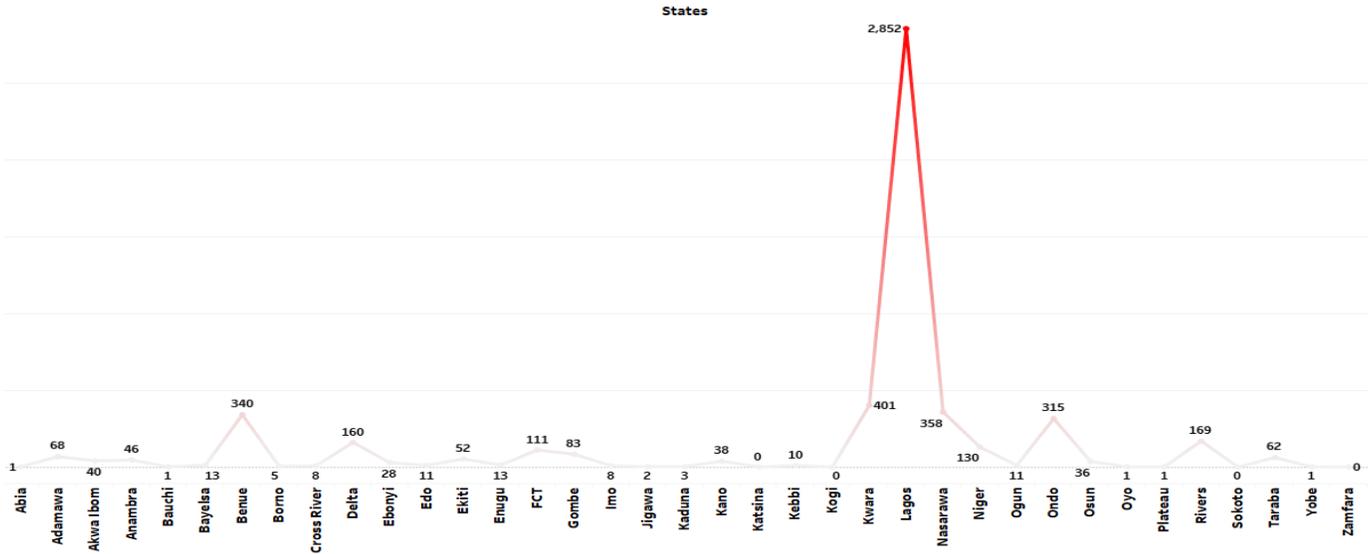

Fig. 3: Total Covid-19 active cases for various states in Nigeria

**3. How many people have recovered and been discharged?**

As of July 19th, 2022, 250,485 people in Nigeria have recovered from Covid-19 and been discharged. The data shows that Lagos, FCT, and Rivers state are the three (3) states with the highest recoveries, while Kebbi, Zamfara, and Kogi state had few recoveries. However, this can also be attributed to their fewer number of confirmed cases.

Fig. 4 below shows a breakdown of the recoveries across the 36 states and FCT:

**4. What is the total number of Covid-19 associated deaths in Nigeria?**

As of July 19th, 2022, Nigeria has recorded 3,144 Covid-19 related deaths. The data shows Lagos, Edo, and FCT are the top three (3) states with Covid-19 related deaths, while Yobe, Zamfara, and Kogi state had the fewest Covid-19 related deaths in Nigeria.

Fig. 5 below shows a breakdown of the deaths across the 36 states and FCT:

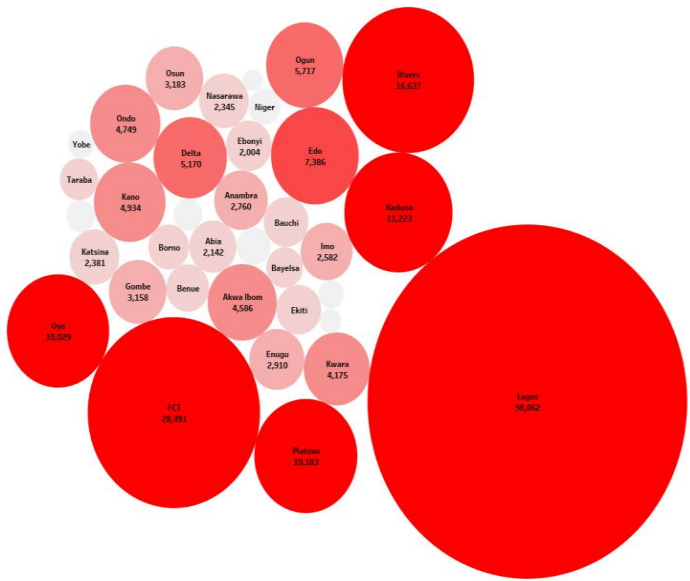

Fig. 4: Total covid-19 recoveries for various states in Nigeria

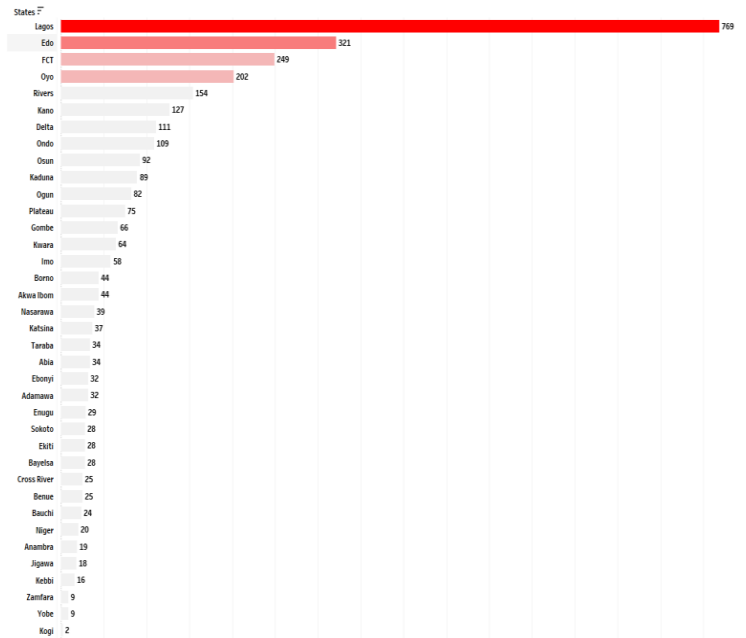

Fig. 5: Deaths associated with covid-19

**5. Dashboard**

Fig. 6 below shows a dashboard of Covid-19 cases across Nigeria between February 28th, 2020, and July 19th, 2022.

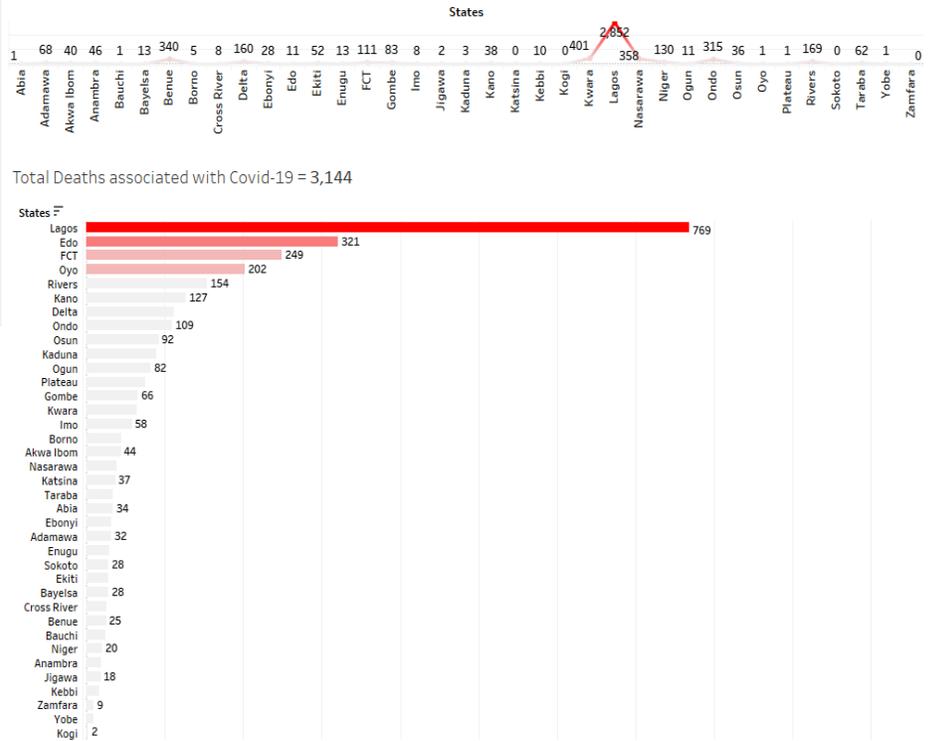

Fig. 6: Dashboard of Covid-19 cases across Nigeria

## V. DISCUSSION & FURTHER INSIGHTS FROM EDA

According to [19] the number of deaths in lab-confirmed positive cases is the most important tool to track the impact of infectious diseases like COVID-19. However, even though the data shows that covid-19 is highly contagious, it still seems less deadly than Ebola, Lassa fever, and yellow fever [15]. We have seen from Fig. 4 vs. Fig. 3 that the rate of recovered cases is very high compared to the active cases, the data also showed that the active cases are very low when compared to the total confirmed cases. Furthermore, the low number of active cases can be traced to some of the positive results of the measures put in place to fight the outbreak in Nigeria especially with increased covid-19 vaccinations as reported by [20]. However, even though the confirmed & active cases and mortality rate may seem low, it is important to note that Nigeria is not out of the woods yet, the virus is still very potent and with Nigeria's elderly population estimated at 6.4 million (people aged > 65 years) [21], the risk of the infection is still there. Also when you take into account other vulnerable populations like those with pre-existing underlying health conditions, there's a need to remain vigilant and continue to monitor the disease and also follow the prevention measures like wearing facemasks and washing of hands.

## VI. CONCLUSION

Today, most academic and industrial research is data-driven; data is also at the center of many modern technological innovations. The enthusiasm for big data analytics and visualization has especially grown in the past few years due to the COVID-19 pandemic sparking people's interest. For many people, it is important to stay informed and understand how Covid-19 cases are progressing and the potential impact on their lives. Also, results from EDA are used by data scientists and analysts from various domains to make better and more informed decisions. This paper presents a visual EDA of the Covid-19 pandemic in Nigeria as reported between February 28th, 2020, and July 19th, 2022. We used data analytics tools (Microsoft Excel and Python libraries) to carry out the data cleaning, preprocessing, and manipulation tasks while the Tableau visualization tool was deployed to create the visuals and dashboard. This project's methodology and results help showcase the effectiveness of big data analytics tools in processing data and presenting insights through visuals/dashboards. These tools can be very effective when dealing with vast amounts of raw and unstructured data.

Finally, bearing in mind the ever-changing nature of current research data records, this study represents a snapshot in time to help future researchers in analyzing the Covid-19 trends in the first 2 years of the pandemic in Nigeria.

## ADDITIONAL INFORMATION

The datasets analyzed and complete documentation of the Exploratory Data Analysis process is available at: https://bit.ly/3qlNqOx. The dashboard and other visualization can be found here: https://tabsoft.co/3u8zety